\documentclass[10pt,twocolumn,letterpaper]{article}

\usepackage{graphicx}
\usepackage{amsmath}
\usepackage{amssymb}
\usepackage{booktabs}

\begin{document}

\title{Weighted Encoding Optimization for Dynamic Single-pixel Imaging and Sensing}

\author{Xinrui Zhan, Liheng Bian, Chunli Zhu, Jun Zhang\\
School of Information and Electronics \& Advanced Research Institute
of Multidisciplinary Science,\\
Beijing Institute of Technology, Beijing, 100081, China\\
{\tt\small bian@bit.edu.cn}
}
\maketitle

\begin{abstract}
  Using single-pixel detection, the end-to-end neural network that jointly optimizes both encoding and decoding enables high-precision imaging and high-level semantic sensing. However, for varied sampling rates, the large-scale network requires retraining that is laboursome and computation-consuming. In this letter, we report a weighted optimization technique for dynamic rate-adaptive single-pixel imaging and sensing, which only needs to train the network for one time that is available for any sampling rates. Specifically, we introduce a novel weighting scheme in the encoding process to characterize different patterns' modulation efficiency. While the network is training at a high sampling rate, the modulation patterns and corresponding weights are updated iteratively, which produces optimal ranked encoding series when converged. In the experimental implementation, the optimal pattern series with the highest weights are employed for light modulation, thus achieving highly-efficient imaging and sensing. The reported strategy saves the additional training of another low-rate network required by the existing dynamic single-pixel networks, which further doubles training efficiency. Experiments on the MNIST dataset validated that once the network is trained with a sampling rate of 1, the average imaging PSNR reaches 23.50 dB at 0.1 sampling rate, and the image-free classification accuracy reaches up to 95.00\% at a sampling rate of 0.03 and 97.91\% at a sampling rate of 0.1.
\end{abstract}

\section{Introduction}

Single-pixel detection is a sensing method that combines image acquisition and information compression together, which effectively saves the storage resources and reduces the cost of processing time \cite{shapiro2008computationa, duarte2008single}.
The single-pixel imaging system is easy-to-operation, which has modest requirements on environment and hardware equipment. A more notable advantage is its sensing ability on non-visible scenarios, which may be unattainable by traditional methods \cite{donoho2006compressed, sun20133d}.
In addition, the superiority further extends the applications of the single-pixel compressive method on encryption and low bandwidth information transmission.
Using the sparse measurements, the imaging and high-level semantic sensing could be enabled by a variety of computing algorithms, in which the representative
traditional solutions include gradient descent, conjugate gradient descent, alternating projection, \emph{etc} \cite{bian2018experimental, edgar2019principles}.
With the significant growth in computing power, deep learning algorithms have also been introduced into the process of single-pixel decoding \cite{sun2016single, higham2018deep}. In particular, for sensing tasks such as classification and segmentation, semantic information can be directly extracted from the sparse measurement without image reconstruction, therefore achieving imaging-free sensing \cite{zhang2020image, liu2021imagefree}.

In the single-pixel encoding process, scenes are compressed to coupled one-dimensional measurements via different optical modulations, namely modulating patterns. Random patterns, Hadamard patterns, Fourier patterns are among the widely used patterns to modulate the target light \cite{hahamovich2021single, zhang2015single, zhang2017fast}. To further improve modulation efficiency, recent studies tend to focus on the optimization of modulating patterns. Sun \emph{et al}. \cite{sun2017russian} designed a sorting method for Hadamard patterns called ``Russian Dolls'' (similar to transform coding), which improves the quality of the reconstructed image at a very low sampling rate. Cao \emph{et al}. \cite{cao2021optimization} proposed an optimal pattern design method inspired by the structure of the biological retina, which improved the imaging quality in the region of interest (ROI). 
However, modulating patterns optimized by the above conventional methods are limited by the complicated and evolving environment. Therefore, it is challenging to obtain the reconstructed image or advanced semantic high-precision sensing results with the poorly targeted modulating patterns. 
To realize efficient imaging and sensing solutions with more realistic scenarios, an end-to-end network has been proposed to optimize the encoding and decoding process jointly \cite{kulkarni2016reconnet, bacca2020coupled, fu2020image}. In a general single-pixel optimization process, the encoding process simulates the modulation step where the modulating patterns are designed as the convolutional layer, while the decoding process represents the decoding network for imaging or sensing. As in \cite{fu2020image}, the proposed end-to-end network is trained collectively for the optimal patterns and the decoding network, which could greatly improve the decoding performance. It is noted that the difference between optimal gray patterns and the actual binary patterns leads to an extra training step for decoding.
To further reduce the network training steps and computational cost for decoding tasks at different sampling rates, a rate-adaptive network, also called as dynamic-rate network, was introduced to single-pixel imaging and sensing \cite{alan2021adaptive, compressed2020xu}. The neural network structure is also an end-to-end network for co-optimizing similar to the work of \cite{kulkarni2016reconnet, bacca2020coupled, fu2020image}, in which dynamic refers to training at a high sampling rate firstly and then retrained the network at a low sampling rate with the fixed decoder network parameters. The two-step training procedure enables the network suitable for tasks with an arbitrarily given sampling rate between the trained sampling rates above.

\section{Method}

\begin{figure}[ht]
\centering
\includegraphics[width=1\linewidth]{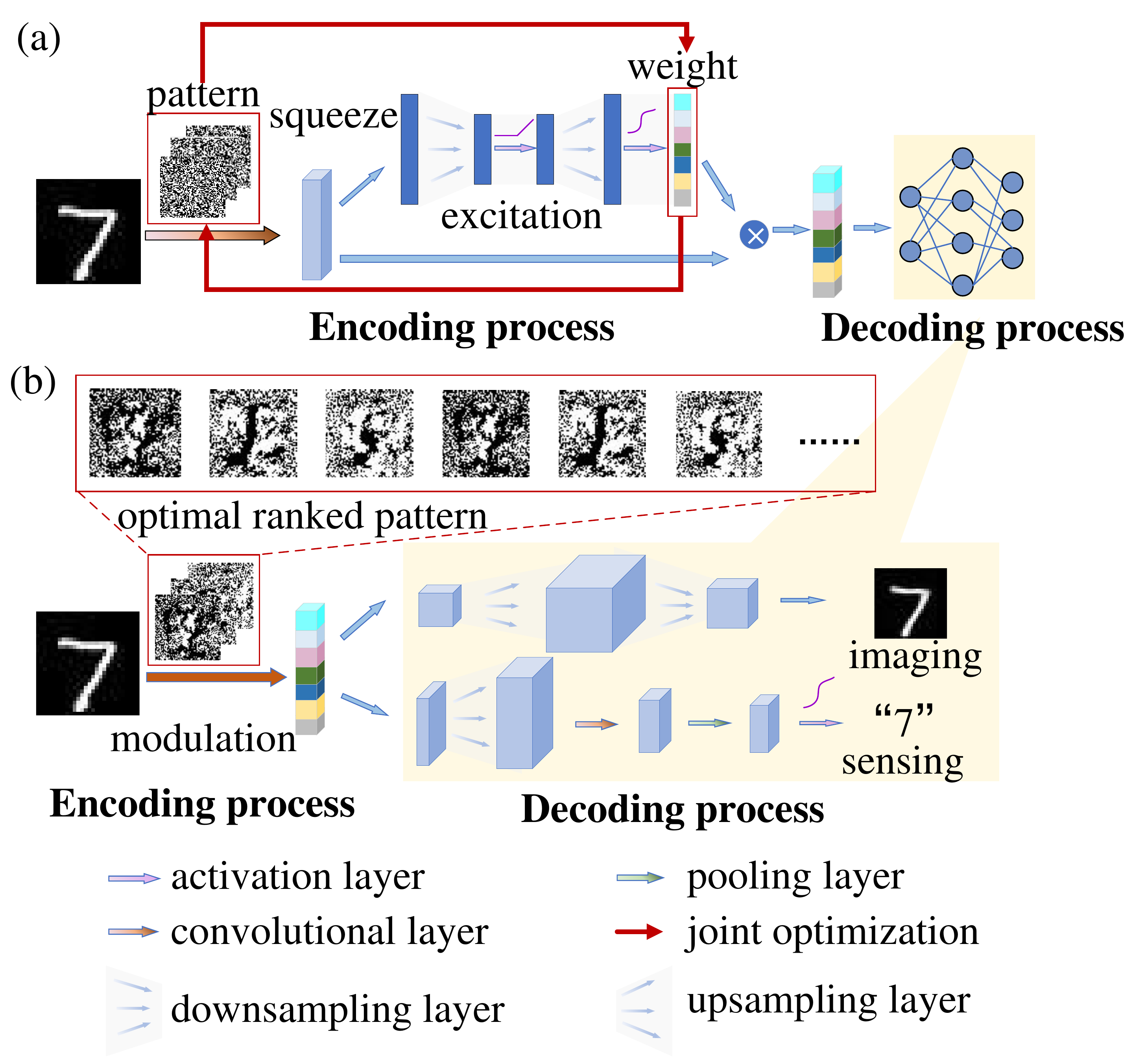}
\caption{Framework of the reported weighted optimization technique for dynamic single-pixel imaging and sensing. (a) The structure of the weighted encoding optimization network. The network is trained for only one time at a high sampling rate to obtain optimal pattern series. (b) The illustration of experimental implementation using the optimal ranked pattern series.}
\label{fig_algrithm}
\end{figure}

In this work, we report a weighted encoding optimization framework for dynamic single-pixel imaging and sensing (as in Fig. \ref{fig_algrithm}). The ``encode-decode'' network architecture with the weighting technique is shown in Fig. \ref{fig_algrithm}(a). In the encoding process, the input image is compressed by a convolutional layer, in which the convolutional kernels and channel numbers represent the physical modulation and modulating frequency, respectively.
Specifically, we introduced a novel weighting architecture based on the SENet structure in the encoding process. SENet \cite{hu2018squeeze} is a neural network based on the attention mechanism, which learns the correlations between channels to calculate their contributions (regarded as weights of patterns' efficiencies in this work) to the network. SENet is composed of two steps: squeezing and excitation. The squeeze operation gathers the global spatial features of each channel into one-dimensional data, while the excitation step learns the dependence on each channel from the low-dimensional data. In the single-pixel modulation experiment, the detection output is a one-dimension vector, so we regard each modulating step as a channel to accomplish the squeeze process. In the excitation operation, we use fully-connected layers and activation functions to measure the characteristic response for different patterns. Then, we obtain the weighted measurement and input it into the decoding network. Eventually, optimal patterns and their appropriate weights can be obtained over several iterations. In the decoding process, the network can be combined with other classical network structures for different purposes. The CNN, VGG and U-Net are commonly used for image reconstruction. For sensing tasks, such as classification and segmentation, an imaging-free strategy can be applied to achieve efficient outputs.

The algorithm of the dynamic single-pixel imaging and sensing with weighted encoding optimization is shown in Fig. \ref{fig_algrithm}. First, we need to train the neural network at a high sampling rate to jointly optimize modulating patterns and their corresponding weights as well as the decoding network. Second, we binarize the optimal patterns using the dithering algorithm \cite{zhang2017fast} and sort them in the order of weight from high to low, as shown in Fig. \ref{fig_algrithm}(b). Practically, for compressive tasks at different sampling rates, the end-to-end network is not required to be trained again for encoding optimization, as shown in Fig. \ref{fig_algrithm}(c). We can directly extract the pattern series from the above step with the highest weights as the optimal ones and correspond these to light modulation. Compared with the existing dynamic rate methods, our method achieved a cost reduction of approximately 50\%, where the network only needs to be trained for one time. Moreover, the reported weighted encoding optimizing strategy merely needs to fine-tune the decoding network via several epochs at any given sampling rate, to obtain a pretrained decoding network. In contrast, existing methods require a new round of joint optimization, thus creating a high computational cost.
Last but not the least, the optimal pattern series is fixed once trained, so for tasks at any sampling rates, it is practicable even when the sampling rate is unclear, which achieves efficient dynamic single-pixel detection.

To verify the effectiveness of our reported weighted modulation optimization, we carried out the following experiments for imaging and sensing tasks, respectively. We use the MNIST dataset \cite{lecun1998mnist} for network training. MNIST is one of the most widely used datasets in deep learning, which consists of handwritten digital images composed of 0 to 9. The dataset contains 60,000 training images and 10,000 test images, in which the sizes of images are unified to 28 $\times$ 28. 

A convolutional decoding network is designed for MNIST image reconstruction. The one-dimensional measurement is expanded into two upsampling layers by deconvolution, and then the feature information is extracted by the pooling operation and activation operation in succession. In the first stage, we train the ``encode-decode'' network at a sampling rate of 1; that is, we set the number of trained patterns in the encoding network as 784. For imaging tasks at any sampling rate smaller than 1, we extract the corresponding binarized series with high weights to be the optimal patterns and then fine-tune the decoding network to achieve efficient imaging. In the experiment, we make comparisons with random patterns, Hadamard patterns, and two-step optimal patterns proposed by Fu \emph{et al}. \cite{fu2020image} at each sampling rate.

\begin{figure}[h]
\centering
\includegraphics[width=1\linewidth]{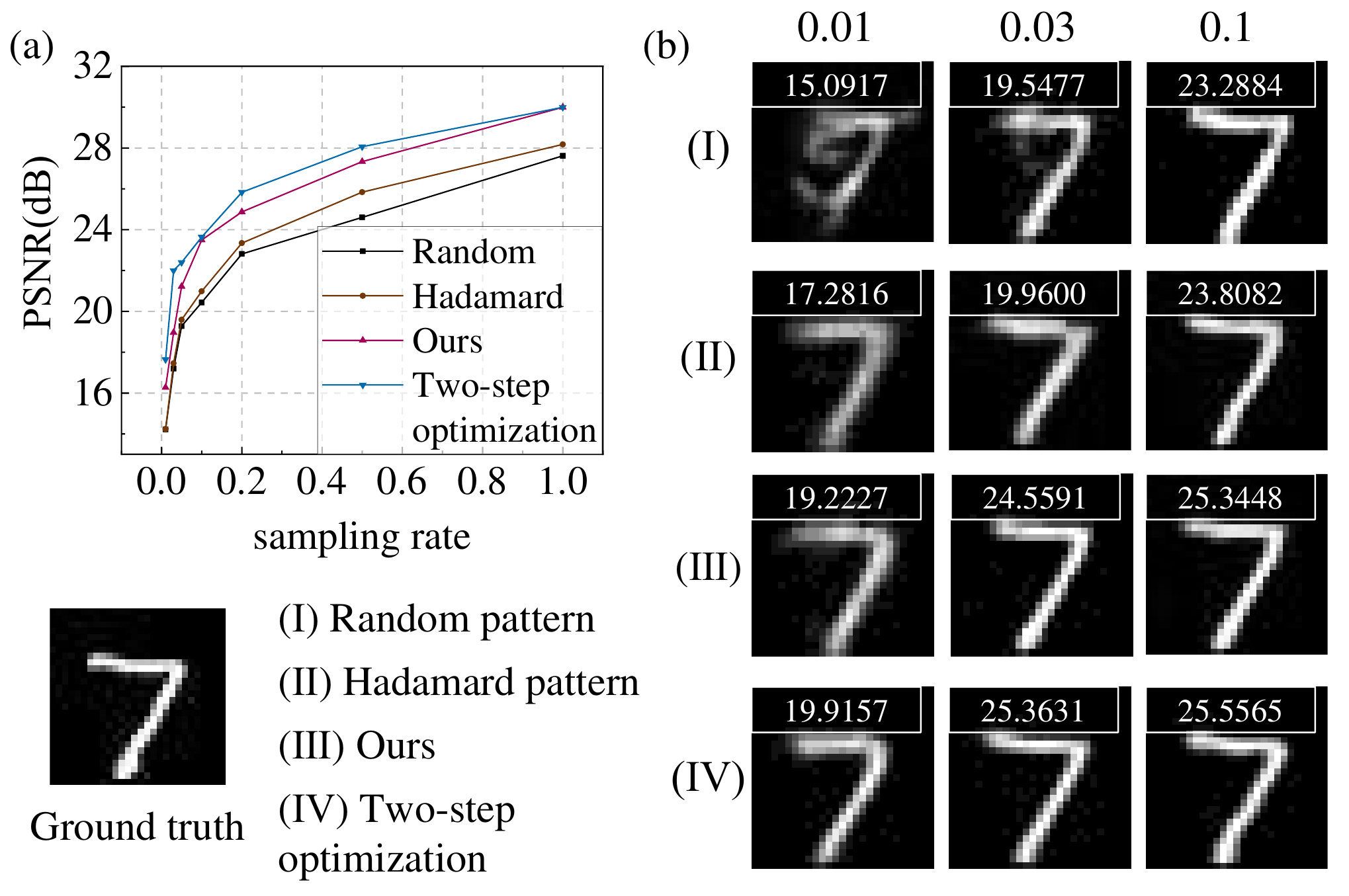}
\caption{Single-pixel imaging results. (a) The average PSNR of the MNIST test dataset by different modulation strategies. (b) Examplar reconstructed images  by different pattern series.}
\label{fig:imaging_result}
\end{figure}

In imaging experiments, we use PSNR(dB) to evaluate the reconstruction results. Fig. \ref{fig:imaging_result}(a) shows the average PSNR of 10,000 testing datasets. It is apparent that at each sampling rate, the PSNR of images reconstructed by the method reported in this work is significantly higher than that of the random patterns and Hadamard patterns, which is higher at least approximately 2 dB. Notably, the substantially reduced computational cost, in fact, does not lead to prominent performance loss.
For example, when the sampling rate is higher than 0.5, the difference gap is lower than 0.7 dB. Fig. \ref{fig:imaging_result}(b) shows the results of a handwritten numeral ``7'' modulated by different pattern series, in which columns from left to right are with the sampling rate of 0.01, 0.03 and 0.1, respectively. At very low sampling rates such as 0.01 and 0.03, the reconstructed image modulated by random patterns or Hadamard patterns is blurred, while ours is clearer. It is generally accepted that the image quality is acceptable when the average PSNR exceeds 24 dB, which sampling rate is 0.5 for traditional modulations and 0.2 for our optimal pattern modulation. 
It can be seen that the optimization reported in this work reduces the sampling rate of single-pixel modulation by approximately 60\% compared with conventional approaches. Meanwhile, the operation steps and calculation consumption have been greatly reduced compared with the two-step method.

In single-pixel sensing experiments, we introduced the imaging-free sensing strategy for MNIST classification. The decoding network is based on EfficientNet \cite{tan2019efficientnet}, which was proposed by Google Researchers in 2019. 
The main target of the EfficientNet is to maximize the model accuracy for any given resource constraints, which uses a neural architecture search strategy to balance the three dimensions of the network: depth, width, and resolution. With the compound scaling method, EfficientNet achieves much better accuracy and efficiency than its former convolutional networks.

\begin{figure}[h]
\centering
\includegraphics[width=1\linewidth]{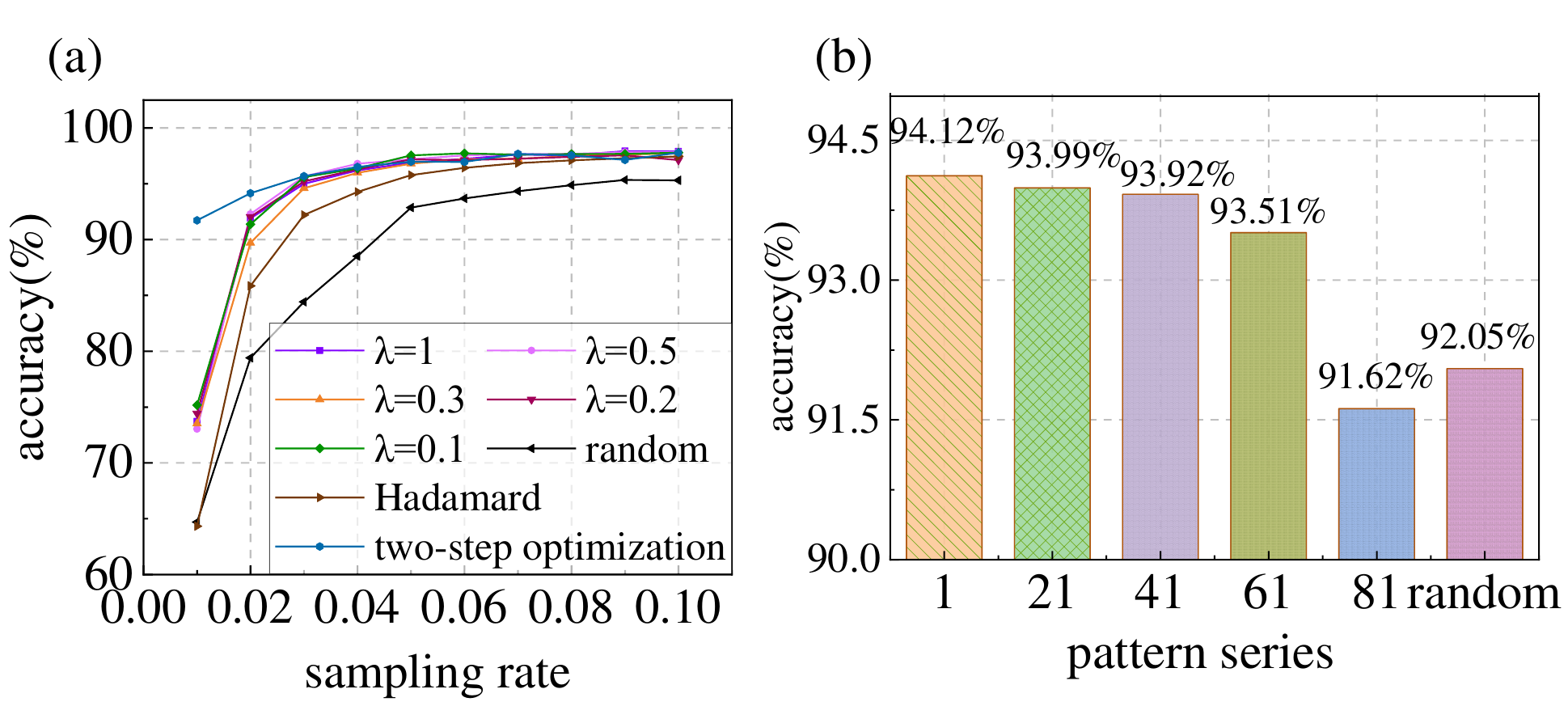}
\caption{Single-pixel sensing results. (a) Classification accuracy of the MNIST test dataset by different modulation strategies. (b) Classification accuracy of the MNIST test dataset by different pattern series of different weights.}
\label{experiment_mnist}
\end{figure}

In the sensing experiment, we compared different modulation methods at each sampling rate, which are weighted optimal patterns, traditional modulation methods (random patterns and Hadamard patterns), and two-step optimal patterns proposed by Fu \emph{et al} \cite{fu2020image}. Before explaining the results, some qualitative analysis about the reported method is discussed as followed.
In the decoding process, we employ a convolutional layer to simulate the light modulation. It is no doubt that the changes of the convolutional layer's channel number will eventually lead to different optimal patterns with distinct values, thus affecting the effectiveness of the subsequent decoding process. In theory, an object's feature information is distributed in the optimal series.
Therefore, when the sampling rate in the physical modulation process is definited, the fewer the number of optimal patterns, the better the performance will be achieved, where feature information is maximally centralized in the selected pattern sequence. However, if we only train a small series of optimal patterns, the ``encode-decode'' network will have a great possibility to be trained again for modulation at higher sampling rates. Essentially, we expected to find a balance between high-performance decoding and enough optimal pattern sequences. To solve this experimentally, in the first step of network training, we set the number of optimal patterns at the sampling rates as 0.1, 0.2, 0.3, 0.5, and 1, and then extracted the corresponding series with the highest weight as the modulating ones at each sampling rate from 0.01 to 0.1.

The result is shown in Fig. \ref{experiment_mnist}(a), where $\lambda$ represents the sampling rate in the first step for training the network. The optimal method achieves higher classification accuracy than traditional methods, especially at low sampling rates. Our method performs better than traditional ones by up to 11.18\%, but not that good at sampling rate less than 0.02 compared with the two-step training. Nevertheless, for sensing tasks at a specific sampling rate of more than 0.03, we can reach almost the same performance as the two-step optimization, which classification accuracy is higher than 94.60\%. Although in theory, higher accuracy may be achieved when patterns are selected from optimal series with a shorter length, sensing result from compressed measurement is not significantly improved. As a general rule, in most single-pixel compressive sensing tasks, the sampling rate is less than 1. Therefore, we recommend optimizing patterns at a sampling rate of 1 in the first step. Then, for tasks at varied sampling rates, there is no need to train the encoding network repeatedly, in which optimal patterns with the highest weights can be directly selected as the modulating patterns to accomplish high-precision sensing.

Furthermore, we performed the following experiment to verify the relevance of the evaluation of the optimal pattern's weight. 
After training the neural network, we sorted the optimal patterns from high to low according to their weights, and then extract different weighted series with the same length to modulate the object. Thereafter, the modulated one-dimensional measurements are trained in the decoding network to realize efficient imaging-free sensing. In the sensing experiment, we regard the classification accuracy as a quantitative indicator to measure the patterns' modulation efficiencies. We optimized 100 patterns in the first step of network training and then selected 20 different patterns for modulation. As shown in Fig. \ref{experiment_mnist}(b), where ``1'' indicates the 1st to the 20th highest weighted patterns, ``21'' indicates the 21st to the 40th highest weighted patterns, and so forth. ``Random'' means 20 patterns were randomly selected regardless of order. The accuracy of images modulated by patterns with low weight is significantly lower than that with higher weight, proving that the weight obtained from SENet is meaningful and effectively measures the pattern's modulation performance. We can also conclude that pattern's modulating efficiency has a great impact on sensing results, where at the same sampling rate of 0.026, the gap of classification accuracy raised by different patterns is up to 2.5\%. 

\section{Experiment}

To experimentally demonstrate the reported method, we built a proof-of-concept setup system and carried out the following experiments. The scene was projected onto the screen by a projector (Panasonic x416c XGA). Afterwards, the reflected light was modulated by a DMD (ViALUX V-7001), and the modulation patterns were the optimal binary series reported in this work. Then modulated light converges through the lens and the coupling signal was collected by a single-pixel silicon photodetector (Thorlabs PDA100A2, 320-1100 nm). By adjusting the frequency of projection and DMD modulation, we could detect a series measurement in the single-pixel detector.

\begin{figure}[h]
\centering
\includegraphics[width=1\linewidth]{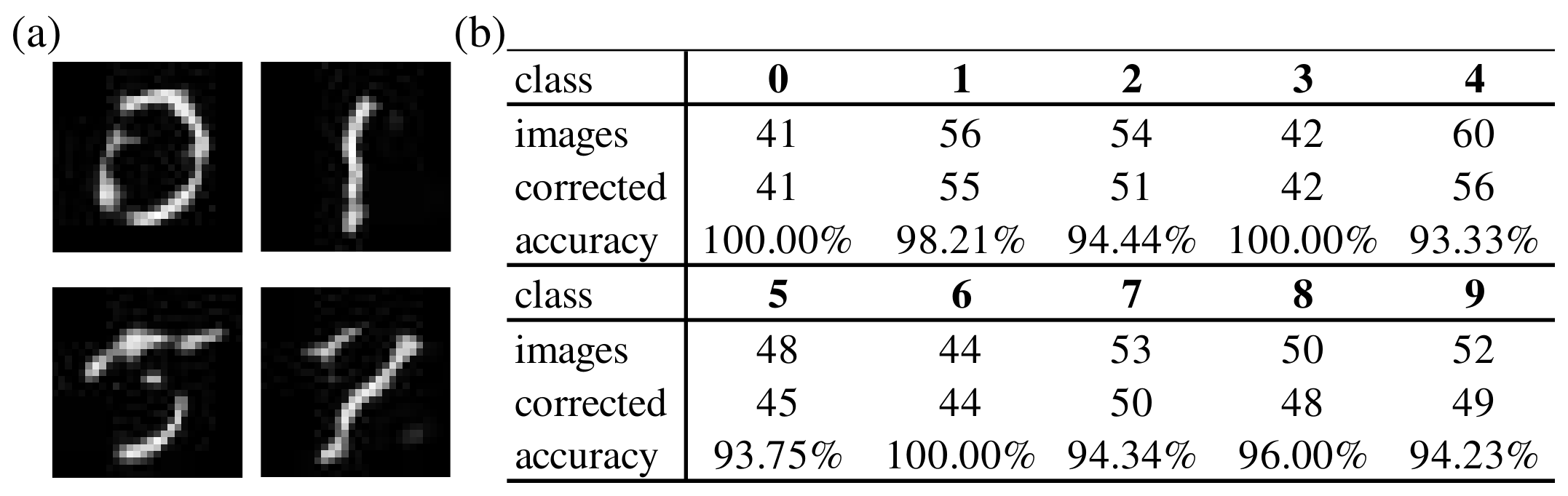}
\caption{Experimental results. (a) Examplar imaging results of the MNIST testing dataset. (b) Classification results for each digit class in the MNIST testing dataset.}
\label{experiment_result}
\end{figure}

The sampling rate in the ``encode-decode'' network was set as 1 to obtain 784 optimal patterns and their weights. For imaging or sensing tasks, we trained the corresponding network separately. Then, the parameters of the encoding part were fixed as the optimal binarized patterns, and the sampling rate was set as 0.1 for both reconstruction and sensing tasks. Specifically, in the encoding process, 78 pattern sequences with the highest weight were selected as fixed parameters in the encoding module; in the physical modulation process, the above optimal patterns were used for modulation. Next, we input the obtained one-dimensional coupling measurement to the pre-trained network to realize efficient imaging and sensing. The reconstruction results are demonstrated in Fig. \ref{experiment_result}(a), which are the imaging results of numbers ``0'', ``1'', ``5'' and ``7''. Results proved that the weighted modulation optimization could reconstruct MNIST images well at a low sampling rate of 0.1. In the sensing experiment, we tested a total of 500 different images. The digit classification result of different number categories are shown in Fig. \ref{experiment_result}(b). We have achieved 100\% accuracy in the recognition of numbers ``0'',`` 3'' and ``6''. 
Although the numbers ``4'', ``5'', ``7'', and ``9'' are with more complicated structures for distinguishing, the accuracy rates all exceed 93\%,
and the overall accuracy reaches up to 96.2\%.

\section{Conclusion}

In this letter, we reported a weighted modulation optimization for dynamic single-pixel imaging and sensing. When optimized at a sampling rate of 1, for the imaging task on the MNIST dataset, the average PSNR of the reconstructed images reaches 23.494 dB at a low sampling rate of 0.1, which is approximately 3 dB better than that of random and Hadamard patterns. On account that we have reduced the computing cost substantially, results revealed almost the same performance with the two-step optimization method (23.642 dB). That comparison further reveals the great potential of the reported method on practical applications, especially in resource-constrained scenarios.
For image sensing, when the sampling rate is set as 0.03, the classification accuracy rate could reach up to 95\%. In essence, the classification accuracy is significantly better than the traditional patterns at each sampling rate. 
In contrast with existing dynamic rate single-pixel methods, our work privileged not only on computation reduction, but also on easier to be conducted and practical on various tasks as the modulating pattern series are fixed regardless of different sampling rates. In addition, our weighted encoding optimization can be applied to achieve efficient dynamic single-pixel detection without any prior knowledge of the sampling rate.

\medskip
\noindent\textbf{Funding.} National Key R\&D Program (Grant No. 2020YFB0505601); National Natural Science Foundation of China (Nos. 62131003, 61827901, 61991451, 61971045).  
\medskip

\noindent\textbf{Disclosures.} The authors declare no conflicts of interest.

\bibliographystyle{refs}
\bibliography{refs}

\begin{thebibliography}{10}
\expandafter\ifx\csname url\endcsname\relax
  \def\url#1{\texttt{#1}}\fi
\expandafter\ifx\csname urlprefix\endcsname\relax\def\urlprefix{URL }\fi
\providecommand{\bibinfo}[2]{#2}
\providecommand{\eprint}[2][]{\url{#2}}

\bibitem{duarte2008single}
\bibinfo{author}{Duarte, M.~F.} \emph{et~al.}
\newblock \bibinfo{title}{Single-pixel imaging via compressive sampling}.
\newblock \emph{\bibinfo{journal}{IEEE Signal Process Mag}}
  \textbf{\bibinfo{volume}{25}}, \bibinfo{pages}{83--91}
  (\bibinfo{year}{2008}).

\bibitem{donoho2006compressed}
\bibinfo{author}{Donoho, D.}
\newblock \bibinfo{title}{Compressed sensing}.
\newblock \emph{\bibinfo{journal}{IEEE Trans. Inf. Theory}}
  \textbf{\bibinfo{volume}{52}}, \bibinfo{pages}{1289--1306}
  (\bibinfo{year}{2006}).

\bibitem{sun20133d}
\bibinfo{author}{Sun, B.} \emph{et~al.}
\newblock \bibinfo{title}{3d computational imaging with single-pixel
  detectors}.
\newblock \emph{\bibinfo{journal}{Science}} \textbf{\bibinfo{volume}{340}},
  \bibinfo{pages}{844--847} (\bibinfo{year}{2013}).
\newblock
  \urlprefix\url{https://www.science.org/doi/abs/10.1126/science.1234454}.
\newblock \eprint{https://www.science.org/doi/pdf/10.1126/science.1234454}.

\bibitem{bian2018experimental}
\bibinfo{author}{Bian, L.}, \bibinfo{author}{Suo, J.}, \bibinfo{author}{Dai,
  Q.} \& \bibinfo{author}{Chen, F.}
\newblock \bibinfo{title}{Experimental comparison of single-pixel imaging
  algorithms}.
\newblock \emph{\bibinfo{journal}{J. Opt. Soc. Am. A}}
  \textbf{\bibinfo{volume}{35}}, \bibinfo{pages}{78--87}
  (\bibinfo{year}{2018}).
\newblock
  \urlprefix\url{http://www.osapublishing.org/josaa/abstract.cfm?URI=josaa-35-1-78}.

\bibitem{edgar2019principles}
\bibinfo{author}{Edgar, M.~P.}, \bibinfo{author}{Gibson, G.~M.} \&
  \bibinfo{author}{Padgett, M.~J.}
\newblock \bibinfo{title}{Principles and prospects for single-pixel imaging}.
\newblock \emph{\bibinfo{journal}{Nature Photon}}
  \textbf{\bibinfo{volume}{13}}, \bibinfo{pages}{13--20}
  (\bibinfo{year}{2019}).

\bibitem{sun2016single}
\bibinfo{author}{Sun, M.-J.} \emph{et~al.}
\newblock \bibinfo{title}{Single-pixel three-dimensional imaging with
  time-based depth resolution}.
\newblock \emph{\bibinfo{journal}{Nat. Commun.}} \textbf{\bibinfo{volume}{7}},
  \bibinfo{pages}{1--6} (\bibinfo{year}{2016}).

\bibitem{higham2018deep}
\bibinfo{author}{Higham, C.~F.}, \bibinfo{author}{Murray-Smith, R.},
  \bibinfo{author}{Padgett, M.~J.} \& \bibinfo{author}{Edgar, M.~P.}
\newblock \bibinfo{title}{Deep learning for real-time single-pixel video}.
\newblock \emph{\bibinfo{journal}{Sci. Rep.}} \textbf{\bibinfo{volume}{8}},
  \bibinfo{pages}{1--9} (\bibinfo{year}{2018}).

\bibitem{zhang2020image}
\bibinfo{author}{Zhang, Z.} \emph{et~al.}
\newblock \bibinfo{title}{Image-free classification of fast-moving objects
  using \&\#x201c;learned\&\#x201d; structured illumination and single-pixel
  detection}.
\newblock \emph{\bibinfo{journal}{Opt. Express}} \textbf{\bibinfo{volume}{28}},
  \bibinfo{pages}{13269--13278} (\bibinfo{year}{2020}).
\newblock
  \urlprefix\url{http://www.osapublishing.org/oe/abstract.cfm?URI=oe-28-9-13269}.

\bibitem{liu2021imagefree}
\bibinfo{author}{Liu, H.}, \bibinfo{author}{Bian, L.} \&
  \bibinfo{author}{Zhang, J.}
\newblock \bibinfo{title}{Image-free single-pixel segmentation}.
\newblock \emph{\bibinfo{journal}{arXiv preprint arXiv:2108.10617}}
  (\bibinfo{year}{2021}).
\newblock \eprint{2108.10617}.

\bibitem{hahamovich2021single}
\bibinfo{author}{Hahamovich, E.}, \bibinfo{author}{Monin, S.},
  \bibinfo{author}{Hazan, Y.} \& \bibinfo{author}{Rosenthal, A.}
\newblock \bibinfo{title}{Single pixel imaging at megahertz switching rates via
  cyclic hadamard masks}.
\newblock \emph{\bibinfo{journal}{Nat. Commun.}} \textbf{\bibinfo{volume}{12}},
  \bibinfo{pages}{1--6} (\bibinfo{year}{2021}).

\bibitem{zhang2015single}
\bibinfo{author}{Zhang, Z.}, \bibinfo{author}{Ma, X.} \&
  \bibinfo{author}{Zhong, J.}
\newblock \bibinfo{title}{Single-pixel imaging by means of fourier spectrum
  acquisition}.
\newblock \emph{\bibinfo{journal}{Nat. Commun.}} \textbf{\bibinfo{volume}{6}},
  \bibinfo{pages}{1--6} (\bibinfo{year}{2015}).

\bibitem{zhang2017fast}
\bibinfo{author}{Zhang, Z.}, \bibinfo{author}{Wang, X.},
  \bibinfo{author}{Zheng, G.} \& \bibinfo{author}{Zhong, J.}
\newblock \bibinfo{title}{Fast fourier single-pixel imaging via binary
  illumination}.
\newblock \emph{\bibinfo{journal}{Sci. Rep.}} \textbf{\bibinfo{volume}{7}},
  \bibinfo{pages}{1--9} (\bibinfo{year}{2017}).

\bibitem{sun2017russian}
\bibinfo{author}{Sun, M.-J.}, \bibinfo{author}{Meng, L.-T.},
  \bibinfo{author}{Edgar, M.~P.}, \bibinfo{author}{Padgett, M.~J.} \&
  \bibinfo{author}{Radwell, N.}
\newblock \bibinfo{title}{A russian dolls ordering of the hadamard basis for
  compressive single-pixel imaging}.
\newblock \emph{\bibinfo{journal}{Sci. Rep.}} \textbf{\bibinfo{volume}{7}},
  \bibinfo{pages}{1--7} (\bibinfo{year}{2017}).

\bibitem{cao2021optimization}
\bibinfo{author}{Cao, J.} \emph{et~al.}
\newblock \bibinfo{title}{Optimization of retina-like illumination patterns in
  ghost imaging}.
\newblock \emph{\bibinfo{journal}{Opt. Express}} \textbf{\bibinfo{volume}{29}},
  \bibinfo{pages}{36813--36827} (\bibinfo{year}{2021}).
\newblock
  \urlprefix\url{http://www.osapublishing.org/oe/abstract.cfm?URI=oe-29-22-36813}.

\bibitem{kulkarni2016reconnet}
\bibinfo{author}{Kulkarni, K.}, \bibinfo{author}{Lohit, S.},
  \bibinfo{author}{Turaga, P.}, \bibinfo{author}{Kerviche, R.} \&
  \bibinfo{author}{Ashok, A.}
\newblock \bibinfo{title}{Reconnet: Non-iterative reconstruction of images from
  compressively sensed measurements}.
\newblock In \emph{\bibinfo{booktitle}{Proceedings of the IEEE Conf. on Comput.
  Vis. Pattern Recognit. (CVPR)}}, \bibinfo{pages}{449--458}
  (\bibinfo{year}{2016}).

\bibitem{bacca2020coupled}
\bibinfo{author}{Bacca, J.}, \bibinfo{author}{Galvis, L.} \&
  \bibinfo{author}{Arguello, H.}
\newblock \bibinfo{title}{Coupled deep learning coded aperture design for
  compressive image classification}.
\newblock \emph{\bibinfo{journal}{Opt. Express}} \textbf{\bibinfo{volume}{28}},
  \bibinfo{pages}{8528--8540} (\bibinfo{year}{2020}).
\newblock
  \urlprefix\url{http://www.osapublishing.org/oe/abstract.cfm?URI=oe-28-6-8528}.

\bibitem{fu2020image}
\bibinfo{author}{Fu, H.}, \bibinfo{author}{Bian, L.} \& \bibinfo{author}{Zhang,
  J.}
\newblock \bibinfo{title}{Single-pixel sensing with optimal binarized
  modulation}.
\newblock \emph{\bibinfo{journal}{Opt. Lett.}} \textbf{\bibinfo{volume}{45}},
  \bibinfo{pages}{3111--3114} (\bibinfo{year}{2020}).
\newblock
  \urlprefix\url{http://www.osapublishing.org/ol/abstract.cfm?URI=ol-45-11-3111}.

\bibitem{alan2021adaptive}
\bibinfo{author}{Yuan, A.~Y.} \emph{et~al.}
\newblock \bibinfo{title}{Adaptive and dynamic ordering of illumination
  patterns with an image dictionary in single-pixel imaging}.
\newblock \emph{\bibinfo{journal}{Opt. Commun.}}
  \textbf{\bibinfo{volume}{481}}, \bibinfo{pages}{126527}
  (\bibinfo{year}{2021}).
\newblock
  \urlprefix\url{https://www.sciencedirect.com/science/article/pii/S0030401820309457}.

\bibitem{compressed2020xu}
\bibinfo{author}{Xu, Y.}, \bibinfo{author}{Liu, W.} \& \bibinfo{author}{Kelly,
  K.~F.}
\newblock \bibinfo{title}{Compressed domain image classification using a
  dynamic-rate neural network}.
\newblock \emph{\bibinfo{journal}{IEEE Access}} \textbf{\bibinfo{volume}{8}},
  \bibinfo{pages}{217711--217722} (\bibinfo{year}{2020}).

\bibitem{hu2018squeeze}
\bibinfo{author}{Hu, J.}, \bibinfo{author}{Shen, L.} \& \bibinfo{author}{Sun,
  G.}
\newblock \bibinfo{title}{Squeeze-and-excitation networks}.
\newblock In \emph{\bibinfo{booktitle}{Proceedings of the IEEE Conf. on Comput.
  Vis. Pattern Recognit. (CVPR)}}, \bibinfo{pages}{7132--7141}
  (\bibinfo{year}{2018}).

\bibitem{lecun1998mnist}
\bibinfo{author}{LeCun, Y.}
\newblock \bibinfo{title}{The mnist database of handwritten digits}.
\newblock \emph{\bibinfo{journal}{http://yann. lecun. com/exdb/mnist/}}
  (\bibinfo{year}{1998}).

\bibitem{tan2019efficientnet}
\bibinfo{author}{Tan, M.} \& \bibinfo{author}{Le, Q.}
\newblock \bibinfo{title}{{E}fficient{N}et: Rethinking model scaling for
  convolutional neural networks}.
\newblock In \bibinfo{editor}{Chaudhuri, K.} \& \bibinfo{editor}{Salakhutdinov,
  R.} (eds.) \emph{\bibinfo{booktitle}{Proceedings of the 36th International
  Conference on Machine Learning}}, vol.~\bibinfo{volume}{97} of
  \emph{\bibinfo{series}{Proceedings of Machine Learning Research}},
  \bibinfo{pages}{6105--6114} (\bibinfo{publisher}{PMLR},
  \bibinfo{year}{2019}).
\newblock \urlprefix\url{https://proceedings.mlr.press/v97/tan19a.html}.

\end{thebibliography}

\end{document}